\documentclass{ws-ijmpcs}

\begin{document}

\markboth{J. Wang}
{Diboson Production at LHC and Tevatron}

%
\catchline{}{}{}{}{}
%

\title{DIBOSON PRODUCTION AT LHC AND TEVATRON}

\author{JIAN WANG\\
on behalf of the ATLAS, CMS, CDF, D\O\ collaborations}

\address{Universite Libre de Bruxelles\\
Brussels, 1050, Belgium\\
Jian.Wang@cern.ch}

\maketitle

\begin{history}
\received{Day Month Year}
\revised{Day Month Year}
\end{history}

\begin{abstract}
This is a report at the conference Physics In Collision 2013. The experimental results on physics of diboson production are reviewed. The measurements use $pp$ collision at the LHC with center-of-mass energy $\sqrt{s}$ = 7 and 8 ~TeV, and $p\bar{p}$ collision at the Tevatron with $\sqrt{s}$ = 1.96~TeV. These include measurements of W$\gamma$, Z$\gamma$, WW, WZ and ZZ production. The results are compared with Standard Model predictions, and are interpreted in terms of constraints on charged and neutral anomalous triple gauge couplings.
\end{abstract}

\ccode{PACS numbers:}

\section{Introduction}	

The study of diboson production provides an important test of the Standard Model (SM) of particle physics at TeV energy scale. Especially, it is sensitive to the self-interactions among vector bosons via triple gauge couplings.
Any significant deviation of the production cross section or kinematic distributions from the SM predictions gives an indication of new physics.
In addition, non-resonant diboson production measurements are important to a precise estimation of irreducible backgrounds for the Higgs study.

The experimental results reviewed here, use $p\bar{p}$ collision at the Tevatron at a center-of-mass energy $\sqrt{s}$ = 1.96~TeV, with an integrated luminosity up to about $10fb^{-1}$, and $pp$ collision at $\sqrt{s}$ = 7 and 8~TeV at the LHC, with integrated luminosities up to about $5~fb^{-1}$ and $20~fb^{-1}$ respectively. Emphasis is placed on the latest results released in the past year.

\section{Cross Section Measurement}

\subsection{W$\gamma$ and Z$\gamma$}

The W$\gamma$ and Z$\gamma$ productions have been studied in $W\gamma \rightarrow l\nu\gamma$ and $Z\gamma \rightarrow ll\gamma$ decay channels at $\sqrt{s}$ = 7~TeV.\cite{Aad:2013izg,Chatrchyan:2013fya} In the W$\gamma$ measurement, events are selected by requiring an isolated electron or muon, and missing transverse energy, $E^{miss}_{T}$, from the undetected neutrino, in addition to an isolated photon. 
The dominant background comes from W+jets, where a jet is misidentified as a photon. In the Z$\gamma$ measurement, events are selected by requiring a same flavor, opposite sign electron or muon pair with an invariant mass close to the Z boson mass, in addition to an isolated photon. The photon is required to be separated from the lepton to suppress the contribution from final state radiation photons.

\begin{figure}[hbtp]
 \begin{center}
   \includegraphics[width=0.45\textwidth]{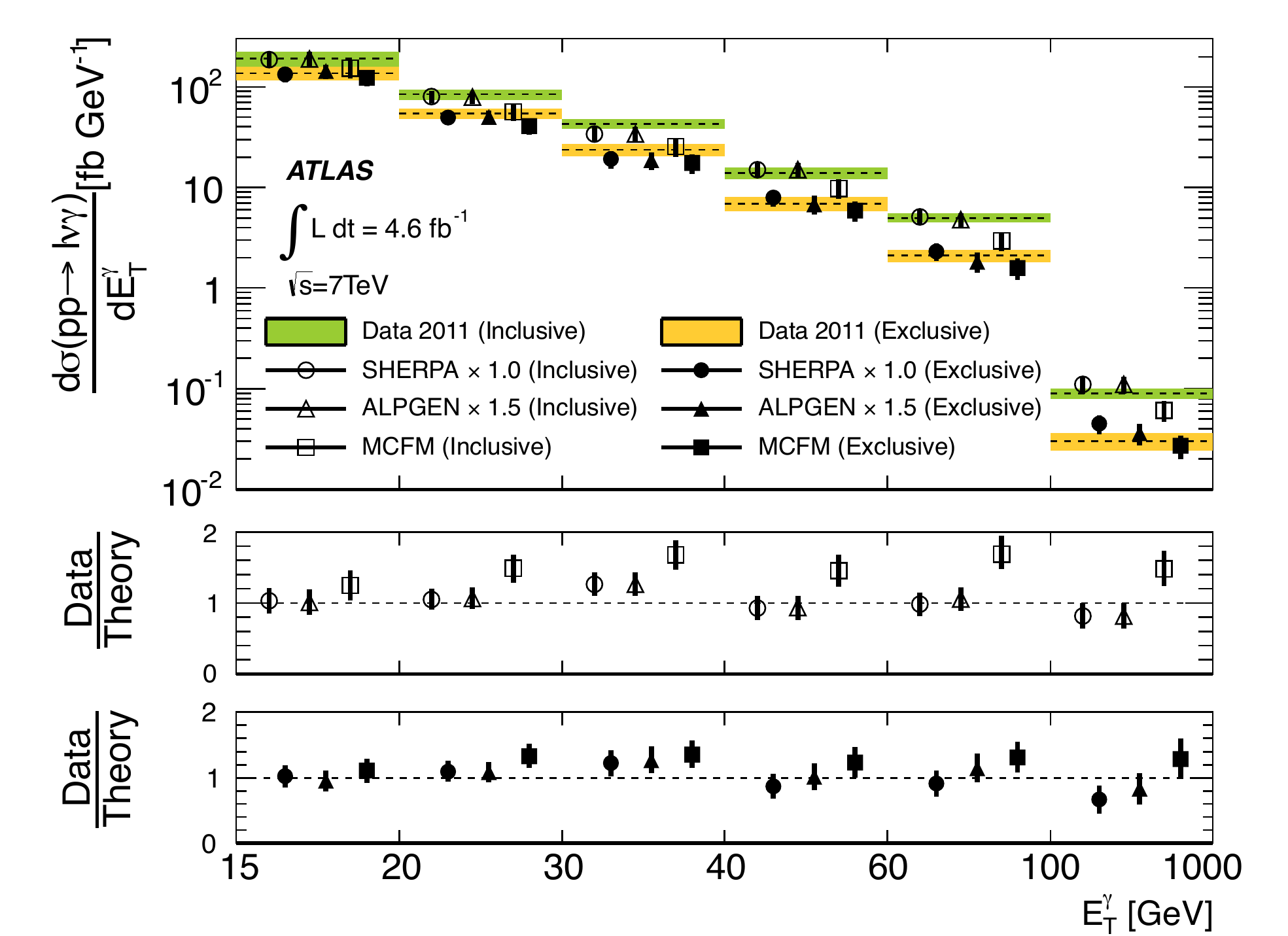}
   \includegraphics[width=0.45\textwidth]{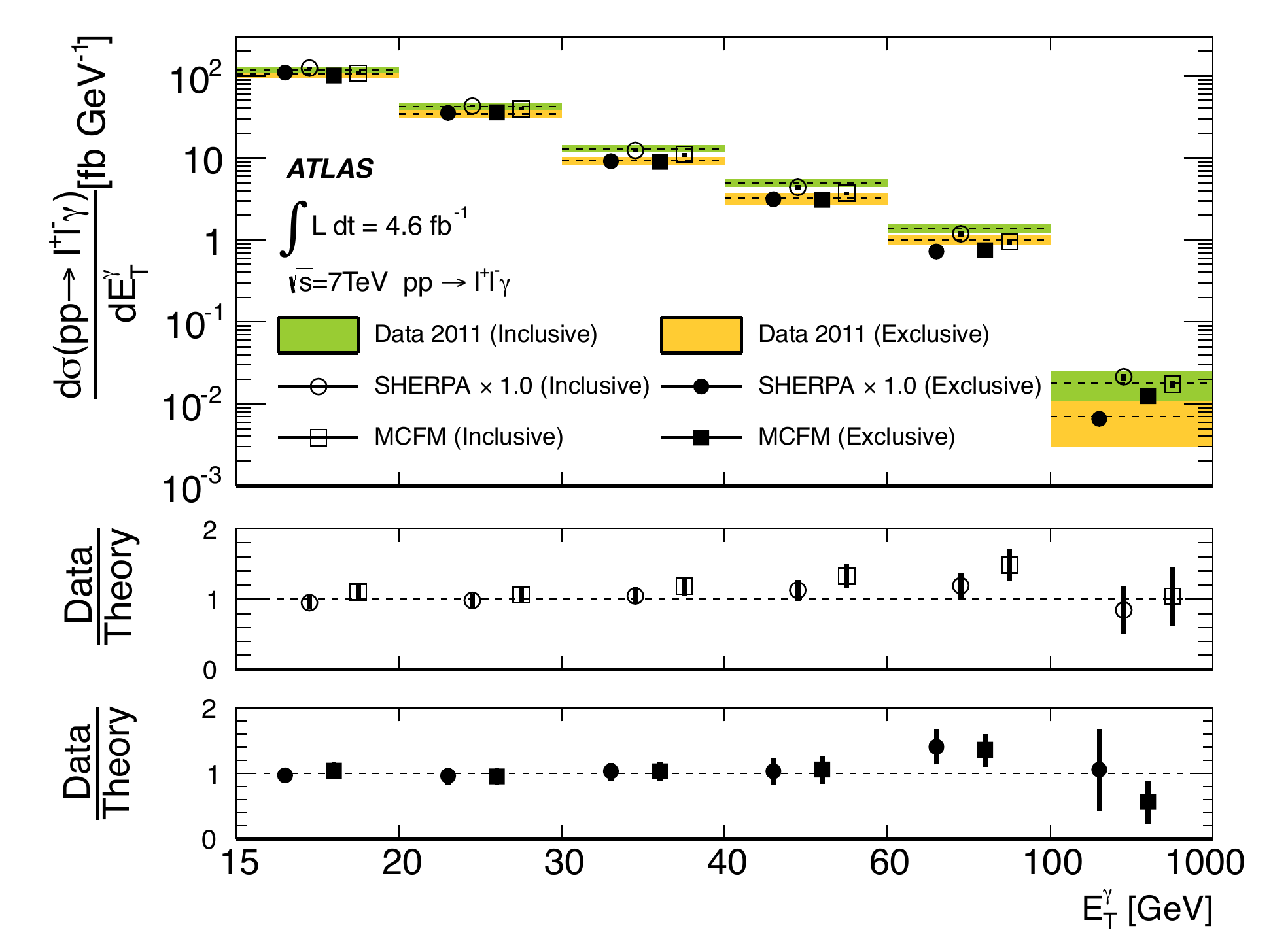}
    \caption{ATLAS measured $\gamma$ $E_{T}$  differential cross sections of the $l\nu\gamma$ process (left) and the $ll\gamma$ process (right), in the inclusive and exclusive (Njet = 0) extended fiducial regions, at $\sqrt{s}$ = 7~TeV}
   \label{fig1}
 \end{center}
\end{figure}

CMS has measured the cross sections for  $\gamma$ $E_{T} > $15~GeV and $m_{ll}>$ 50~GeV. The measured W$\gamma$ cross section times $W \rightarrow l\nu$ branching ratio is $37.0 \pm 0.8~(stat.) \pm 4.0~(syst.) \pm 0.8~(lumi.)pb$. The measured Z$\gamma$ cross section times $Z \rightarrow ll$ branching ratio is $5.33 \pm 0.08~(stat.) \pm 0.25~(syst.) \pm 0.12~(lumi.)pb$. The results are consistent with the SM predictions.
The differential cross sections measured by ATLAS, comparing with theoretical predictions, are shown in Figure~\ref{fig1}. In general, the NLO parton-level MC, MCFM,\cite{Campbell:2010ff,Campbell:2011bn} agrees with the exclusive (Njet = 0) production cross section measurements, while LO MC (ALPGEN\cite{Mangano:2002ea} or SHERPA\cite{Gleisberg:2008ta}) with multiple parton emission reproduce the $\gamma$ $E_{T}$ spectrum.

The $Z\gamma \rightarrow \nu\nu\gamma$ decay channel has also been measured.\cite{Aad:2013izg,Chatrchyan:2013nda} The backgrounds are jets misidentified as photons, and instrumental sources such as beam-gas interactions. 
Very tight photon $E_{T}$ and $E^{miss}_{T}$ cuts are applied to suppress such backgrounds. Detector timing is also used to reduce instrumental backgrounds. Both ATLAS and CMS results are in agreement with the SM predictions.
 
\subsection{WW}

The WW production cross section has been measured in the $WW \rightarrow l\nu l'\nu'$ final state.\cite{ATLAS:2012mec,Chatrchyan:2013yaa,Chatrchyan:2013oev} Events are selected by requiring two opposite charged isolated leptons, electron or muon, accompanied by significant $E^{miss}_{T}$.
The Z+jets background in the $ee$ and $\mu\mu$ channels is suppressed by a cut on large $E^{miss}_{T}$ and a Z veto. To minimise the contribution from top-quark background, events containing jets are rejected (jet veto). This leads to a significant theoretical uncertainty in the jet veto efficiency.

\begin{figure}[hbtp]
 \begin{center}
   \includegraphics[width=0.55\textwidth]{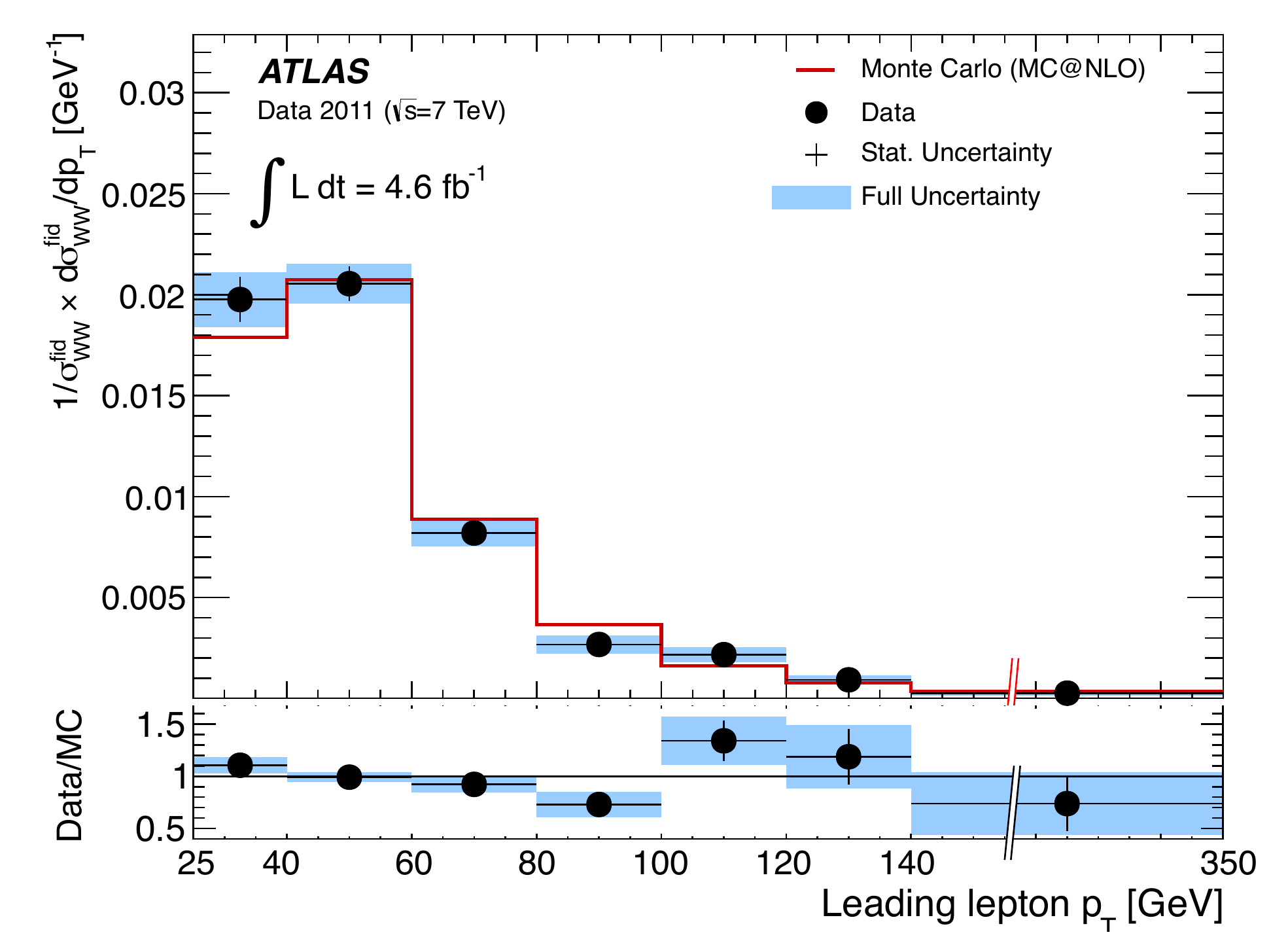}
    \caption{ATLAS measured normalized differential WW fiducial cross section as a function of the leading lepton $p_{T}$ compared to the SM prediciton.}
   \label{fig5}
 \end{center}
\end{figure}

The WW production cross sections measured by ATLAS and CMS at $\sqrt{s}$ = 7~TeV are $51.9 \pm 2.0~(\rm{stat.})\pm 3.9~(\rm{syst.})\pm 2.0~(\rm{lumi.})pb$ and $52.4 \pm 2.0~(\rm{stat.})\pm 4.5~(\rm{syst.})\pm 1.2~(\rm{lumi.})pb$ respectively.  The ATLAS measurement of differential cross section as a fuction of the leading lepton $p_{T}$ is shown in Figure~\ref{fig5}. 
CMS has also performed a first measurement of WW production at $\sqrt{s}$ = 8~TeV. The cross sections is determined to be $69.9 \pm 2.8~(\rm{stat.})\pm 5.6~(\rm{syst.})\pm 3.1~(\rm{lumi.})pb$. These measured cross sections are slightly higher than but still compatible with the SM predictions. Systematic uncertainties dominate the total uncertainty.

\subsection{WZ}

The WZ production cross sections have been measured in the $WZ \rightarrow l\nu 2l'$ decay channel.\cite{Aad:2012twa,ATLAS:2013fma,CMS:2013qea} This final state has very low background after requiring exactly three isolated leptons (electron or muon), a pair of which is of same-flavor and has an invariant mass close to the mass of Z boson, in addition to significant $E^{miss}_{T}$ from W decay. 

In ATLAS measurement at $\sqrt{s}$ = 7~TeV, totally 317 candidates are observed with a background expectation of 68 events. 
In the measurement at $\sqrt{s}$ = 8~TeV, 1094 candidate events are observed in total, with a background expectation of 227 events.
The WZ cross sections are measured to be $19.0^{+1.4}_{-1.3}~(\rm{stat.})\pm 0.9~(\rm{syst.})\pm 0.4~(\rm{lumi.})pb$ at $\sqrt{s}$ = 7~TeV, and  $20.3^{+0.8}_{-0.7}~(\rm{stat.})^{+1.2}_{-1.1}~(\rm{syst.})^{+0.7}_{-0.6}~(\rm{lumi.})pb$ at $\sqrt{s}$ = 8~TeV, for the Z boson mass in the range of 66 to 116~GeV.

CMS has measured the WZ production cross section for the Z boson mass between 71 to 111~GeV. The cross sections are determined to be $20.76 \pm 1.32~(\rm{stat.})\pm 1.13~(\rm{syst.})\pm 0.46~(\rm{lumi.})pb$ at $\sqrt{s}$ = 7~TeV, and $24.61 \pm 0.76~(\rm{stat.})\pm 1.13~(\rm{syst.})\pm 1.08~(\rm{lumi.})pb$ at $\sqrt{s}$ = 8~TeV. Since the LHC is a $pp$ collider, the $W^{+}Z$ and $W^{-}Z$ cross sections are not equal. The ratios of production cross sections for $W^{+}Z$ and $W^{-}Z$ have also been measured. They are $1.94\pm 0.25~(\rm{stat.})\pm 0.04~(\rm{syst.})$ and $1.81\pm 0.12~(\rm{stat.})\pm 0.03~(\rm{syst.})$ at $\sqrt{s}$ = 7~TeV and 8~TeV respectively, in agreement with the SM predictions.

The total production cross section of WV (WW+WZ) has also been studied in the $WV \rightarrow l\nu qq$ final state at $\sqrt{s}$ = 7~TeV.\cite{ATLAS:2012dtt,Chatrchyan:2012bd} The resolution of reconstructed di-jet mass is about 10~GeV, which cannot distinguish W from Z here. This channel has higher branching ratio with respect to the fully leptonic decay mode, at the cost of larger W/Z+jets background. Events are selected by requiring an isolated electron or muon, $E^{miss}_{T}$ and exactly two high-$p_{T}$ jets. Signal is extracted by fitting di-jet mass distribution. ATLAS measures a cross section of $72 \pm 9~(\rm{stat.})\pm 15~(\rm{syst.})\pm 13~(\rm{MC~stat.})pb$ and CMS measures a cross section of $68.9 \pm 8.7~(\rm{stat.})\pm 9.7~(\rm{syst.})\pm 1.5~(\rm{lumi.})pb$. Both measurements agree with the SM predictions.

\subsection{ZZ}

The ZZ production cross sections have been measured using the high purity four-lepton ($ZZ \rightarrow 2l2l'$) decay channel.\cite{Aad:2012awa,ATLAS:2013gma,Chatrchyan:2012sga,CMS:2013hea}. Even though the branching ratio to four-lepton final state is small, this process is really clean, with negligible amount of background. Events are selected by requiring two pairs of electrons or muons, with opposite-charge and same-flavour. The invariant mass of each pair is compatible with the Z boson mass. The challenge in the four-lepton analysis is the optimization for lepton efficiencies, especially for low $p_{T}$ leptons. Some leptons might fall outside the acceptance of the detector while some others may fail the criteria used to select a lepton. With four chances to miss a lepton, even small inefficiencies will add up.

In ATLAS measurement at $\sqrt{s}$ = 7~TeV, $ZZ \rightarrow 2l2\nu$ decay mode has also been measured, by applying a tight cut on a $E^{miss}_{T}$-related variable in order to suppress the dominant Z+jets background. $2l2l'$ and $2l2\nu$ results are then combined, assuming the SM branching ratios. The ZZ production cross section is determined to be $6.7\pm 0.7~(\rm{stat.})^{+0.4}_{-0.3}~(\rm{syst.})\pm 0.3~(\rm{lumi.})pb$ at $\sqrt{s}$ = 7~TeV, where both Z bosons in the mass range of 66 to 116~GeV.
In the measurement at $\sqrt{s}$ = 8~TeV, only four-lepton channel is used. Totally 305 candidate events are observed with a background expectation of 20.4. The SM expectation for the number of signal event is 292.5. The ZZ production cross section  at $\sqrt{s}$ = 8~TeV is measured to be $7.1^{+0.5}_{-0.4}~(\rm{stat.})\pm 0.3~(\rm{syst.})\pm 0.2~(\rm{lumi.})pb$.
 
\begin{figure}[hbtp]
 \begin{center}
   \includegraphics[width=0.45\textwidth]{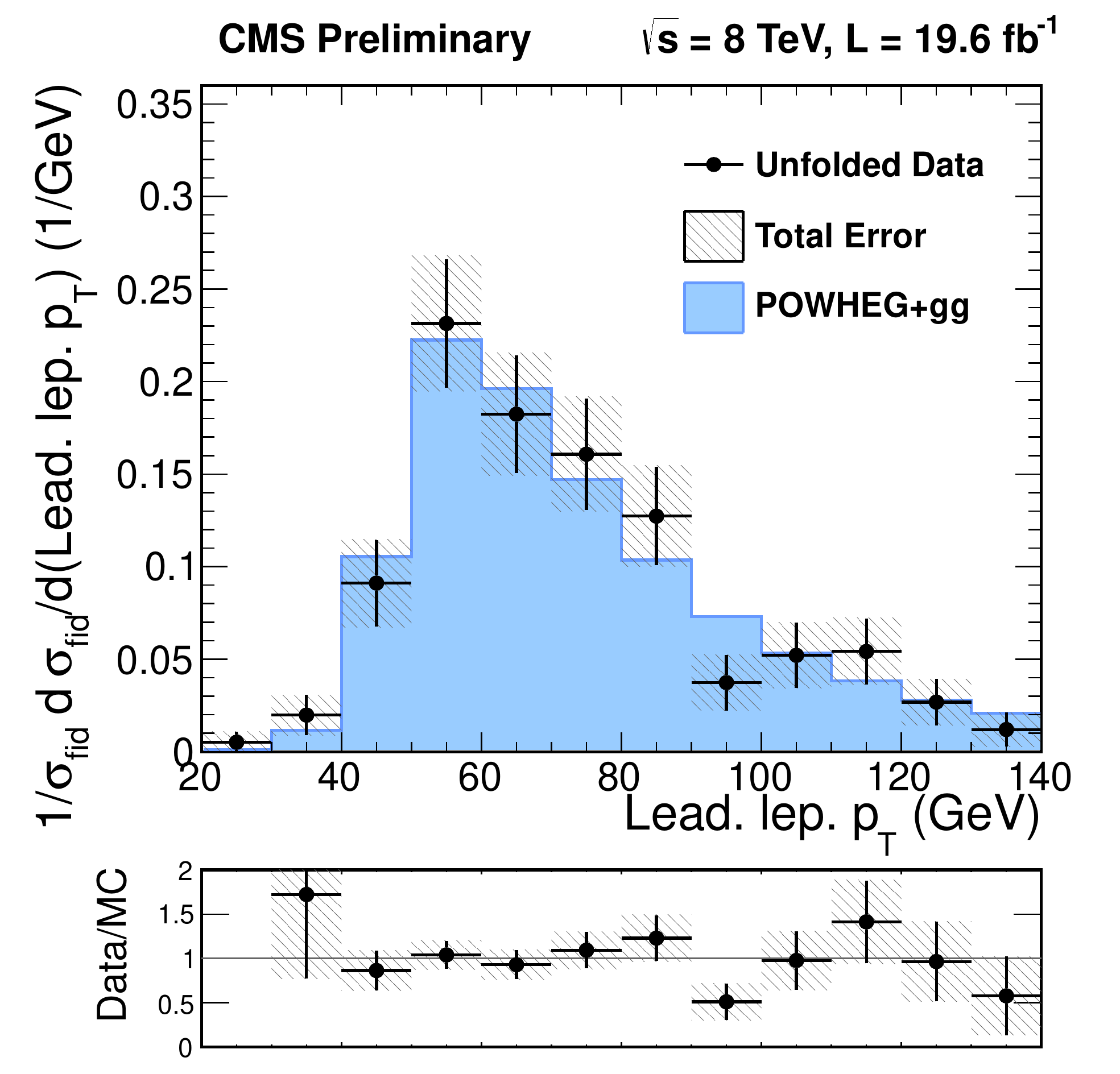}
   \includegraphics[width=0.45\textwidth]{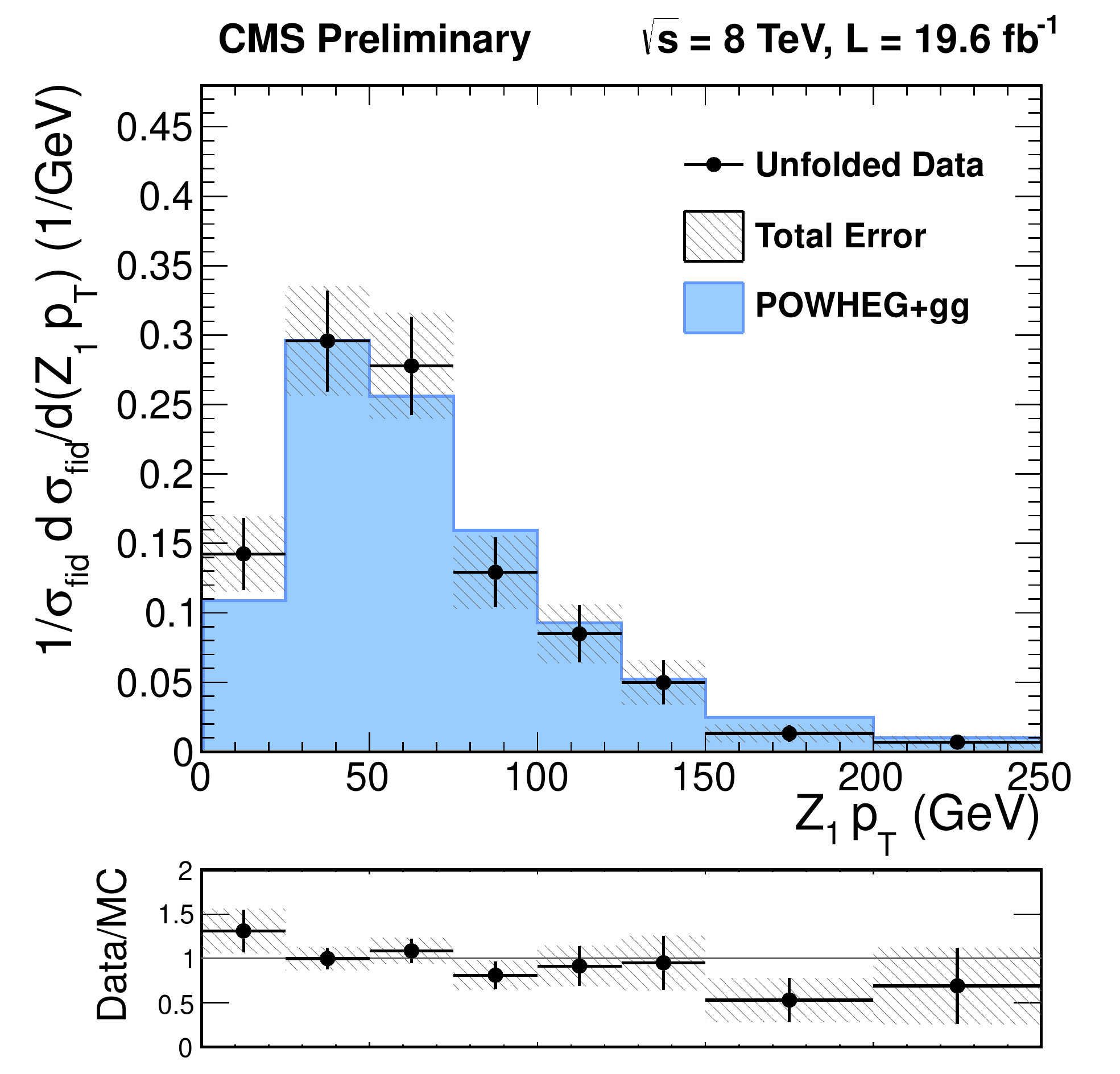}
    \caption{Differential cross section normalized to the fiducial cross section in the CMS $ZZ \rightarrow 2l2l'$ measurement at $\sqrt{s}$ = 8~TeV. The differential cross sections in bins of $p_{T}$ are presented for the leading lepton (left) and the higher-$p_{T}$ Z (right).}
   \label{fig2}
 \end{center}
\end{figure}

The CMS $ZZ \rightarrow 2l2l'$ measurements include $ZZ \rightarrow 2l2\tau$ decay mode as well. The measured cross sections are $6.24^{+0.86}_{-0.80}~(\rm{stat.})^{+0.41}_{-0.32}~(\rm{syst.})\pm 0.14~(\rm{lumi.})pb$ at $\sqrt{s}$ = 7~TeV and $7.7^{+0.5}_{-0.5}~(\rm{stat.})^{+0.5}_{-0.4}~(\rm{syst.})\pm 0.4~(\rm{theo.}) \pm 0.3~(\rm{lumi.})pb$ at $\sqrt{s}$ = 8~TeV, for both Z bosons produced in the mass region of 60 to 120~GeV. Differential cross sections are also measured (as shown in Figure~\ref{fig2}) and well described by the theoretical predictions.

CDF and D\O\ have measured ZZ cross section in $p\bar{p}$ collision at $\sqrt{s}$ = 1.96~TeV.\cite{D0:2013rca} CDF has studied ZZ production through $2l2l'$ and $2l2\nu$ final states, using a dataset corresponding to 9.7$fb^{-1}$ integrated luminosity. The combined measured ZZ cross section is $1.04^{+0.20}_{-0.24}~(\rm{stat.})^{+0.15}_{-0.08}~(\rm{syst.})$.
D\O\ has measured the four-lepton channel, and combined it with a previous study of $2l2\nu$ channel, resulting a ZZ cross section of  $1.32^{+0.29}_{-0.25}~(\rm{stat.})\pm 0.12~(\rm{syst.})\pm 0.04~(\rm{lumi.})$ .

A summary of WW, WZ and ZZ production cross section measurements are listed and compared with the relevant theoretical predictions in Table~\ref{tab1}. The theoretical predictions are computed using MCFM to QCD NLO. Please note the theoretical predictions are different for same process measured by different experiments, because the phase spaces are not exactly same defined.

\begin{table}[ph]
\tbl{Summary of diboson production cross section measurements. }
{\begin{tabular}{@{}ccccc@{}} \toprule
Experiment & $\sqrt{s}$ & Integrated luminosity & Measured cross section & Theoretical prediction \\
& (~TeV) &  ($fb^{-1}$) & (pb) & (pb) \\ \colrule 
WW \\ \botrule
ATLAS & 7 & 4.6 & $51.9 \pm 2.0~(\rm{stat.})\pm 3.9~(\rm{syst.})\pm 2.0~(\rm{lumi.})$ & $44.7^{+2.1}_{-1.9} $\\  \colrule
CMS & 7 & 4.9 &  $52.4 \pm 2.0~(\rm{stat.})\pm 4.5~(\rm{syst.})\pm 1.2~(\rm{lumi.})$ & $47.0\pm 2.0$\\ \colrule
CMS & 8 & 3.5 & $69.9 \pm 2.8~(\rm{stat.})\pm 5.6~(\rm{syst.})\pm 3.1~(\rm{lumi.})$ & $57.3^{+2.3}_{-1.6} $\\ \botrule
WZ \\ \botrule
ATLAS & 7 & 4.6 & $19.0^{+1.4}_{-1.3}~(\rm{stat.})\pm 0.9~(\rm{syst.})\pm 0.4~(\rm{lumi.})$ & $17.6^{+1.1}_{-1.0} $\\ \colrule
ATLAS & 8 &  13 & $20.3^{+0.8}_{-0.7}~(\rm{stat.})^{+1.2}_{-1.1}~(\rm{syst.})^{+0.7}_{-0.6}~(\rm{lumi.})$ & $20.3\pm 0.8 $\\ \colrule
CMS & 7 & 4.9 & $20.76 \pm 1.32~(\rm{stat.})\pm 1.13~(\rm{syst.})\pm 0.46~(\rm{lumi.})$ & $17.8^{+0.7}_{-0.5} $\\ \colrule
CMS & 8 & 19.6 & $24.61 \pm 0.76~(\rm{stat.})\pm 1.13~(\rm{syst.})\pm 1.08~(\rm{lumi.})$ & $21.91^{+1.17}_{-0.88} $\\ \botrule
WV (V= W or Z) \\ \colrule
ATLAS & 7 & 4.7 & $72 \pm 9~(\rm{stat.})\pm 15~(\rm{syst.})\pm 13~(\rm{MC~stat.})$ & $63.4 \pm 2.6 $\\ \colrule
CMS & 7 & 5 & $68.9 \pm 8.7~(\rm{stat.})\pm 9.7~(\rm{syst.})\pm 1.5~(\rm{lumi.})$ & $65.6 \pm 2.2 $\\ \botrule
ZZ  \\ \botrule
ATLAS & 7 & 4.6 & $6.7\pm 0.7~(\rm{stat.})^{+0.4}_{-0.3}~(\rm{syst.})\pm 0.3~(\rm{lumi.})$ & $5.89^{+0.22}_{-0.18} $\\ \colrule
ATLAS & 8 & 20 & $7.1^{+0.5}_{-0.4}~(\rm{stat.})\pm 0.3~(\rm{syst.})\pm 0.2~(\rm{lumi.})$ & $7.2^{+0.3}_{-0.2} $\\ \colrule
CMS & 7 & 5.0 & $6.24^{+0.86}_{-0.80}~(\rm{stat.})^{+0.41}_{-0.32}~(\rm{syst.})\pm 0.14~(\rm{lumi.})$ & $6.3\pm 0.4$\\ \colrule
CMS & 8 &  19.6 & $7.7^{+0.5}_{-0.5}~(\rm{stat.})^{+0.5}_{-0.4}~(\rm{syst.})\pm 0.4~(\rm{theo.}) \pm 0.3~(\rm{lumi.})$ & $7.7\pm 0.6$\\ \colrule 
CDF & 1.96($p\bar{p}$) & 9.7 & $1.04^{+0.20}_{-0.24}~(\rm{stat.})^{+0.15}_{-0.08}~(\rm{syst.})$ & $1.4 \pm 0.1 $\\ \colrule
D\O\ & 1.96($p\bar{p}$) & 9.8 & $1.32^{+0.29}_{-0.25}~(\rm{stat.})\pm 0.12~(\rm{syst.})\pm 0.04~(\rm{lumi.})$ & $1.43\pm 0.10 $\\ \botrule
\end{tabular} \label{tab1}}
\end{table}

\section{Limits On Anomalous Triple Gauge Couplings}

The SM describes exactly how vector bosons couple with each other, and diboson productions are sensitive to these couplings. Even if new physics is at very high energy scale, beyond the reach of current colliders (which means direct pair production of new particle is impossible ), it could still have indirect effect on triple gauge couplings through virtual corrections.
Anomalous triple gauge couplings (aTGC) could be modeled by adding terms to the SM Lagrangian, using a set of parameters, listed in Table~\ref{tab2}. All these parameters are equal to zero in the SM. This is a common approach to parameterize low energy effects from high energy scale new physics, which allows for experimental results to be interpreted as model independent constrains on aTGC. 

\begin{table}[ph]
\tbl{Parameterization of aTGC.}
{\begin{tabular}{@{}ccc@{}} \toprule
Coupling & Parameters & Channels \\ \colrule
WW$\gamma$ & $\Delta \kappa_{\gamma}, \lambda_{\gamma}$ & WW, W$\gamma$ \\ \colrule
WWZ & $\Delta g^{Z}_{1}, \Delta \kappa_{Z}, \lambda_{Z}$ & WW,WZ \\ \colrule
ZZ$\gamma$ & $h^{Z}_{3}, h^{Z}_{4}$ & Z$\gamma$ \\ \colrule
Z$\gamma\gamma$ & $h^{\gamma}_{3}, h^{\gamma}_{4}$ & Z$\gamma$ \\ \colrule
ZZZ & $ f^{Z}_{4}, f^{Z}_{5} $ & ZZ \\ \colrule
Z$\gamma$Z  & $ f^{\gamma}_{4}, f^{\gamma}_{5} $ & ZZ \\ \botrule
\end{tabular} \label{tab2}}
\end{table}

\begin{figure}[hbtp]
 \begin{center}
   \includegraphics[width=0.65\textwidth]{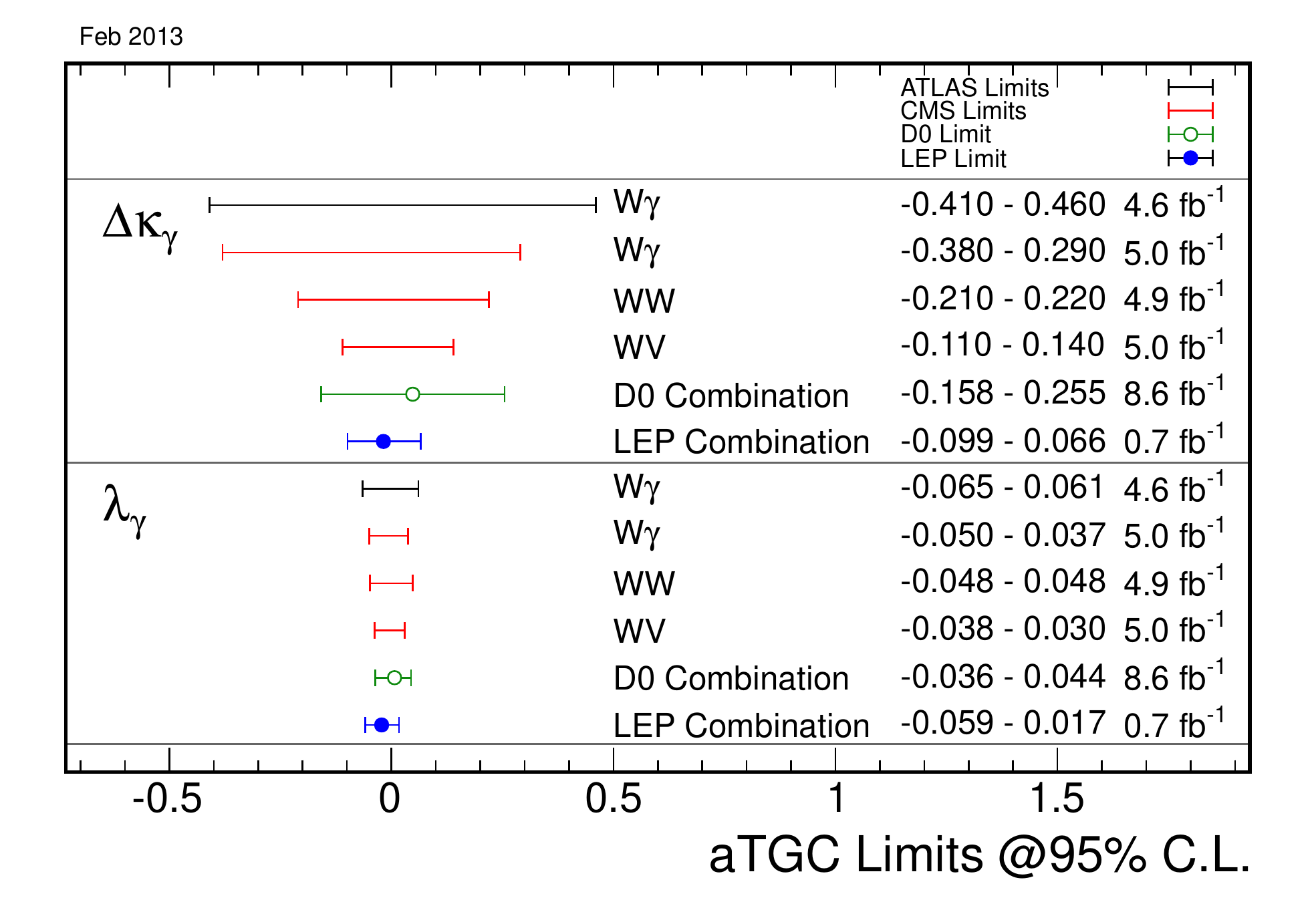}
   \includegraphics[width=0.65\textwidth]{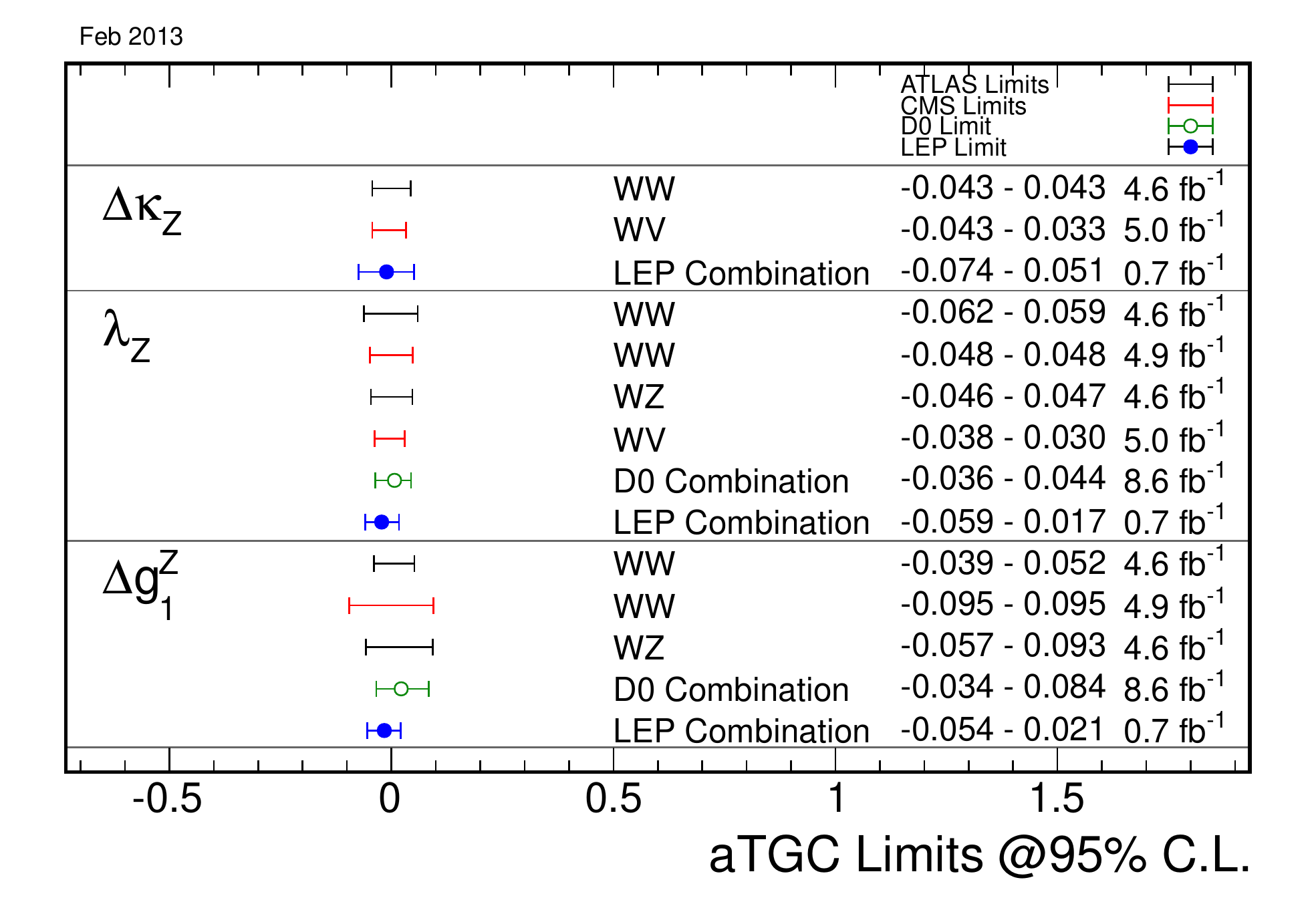}
    \caption{Limits at 95\% C.L. on WW$\gamma$ (top) and WWZ (bottom) aTGC.}
   \label{fig3}
 \end{center}
\end{figure}

\begin{figure}[hbtp]
 \begin{center}
   \includegraphics[width=0.65\textwidth]{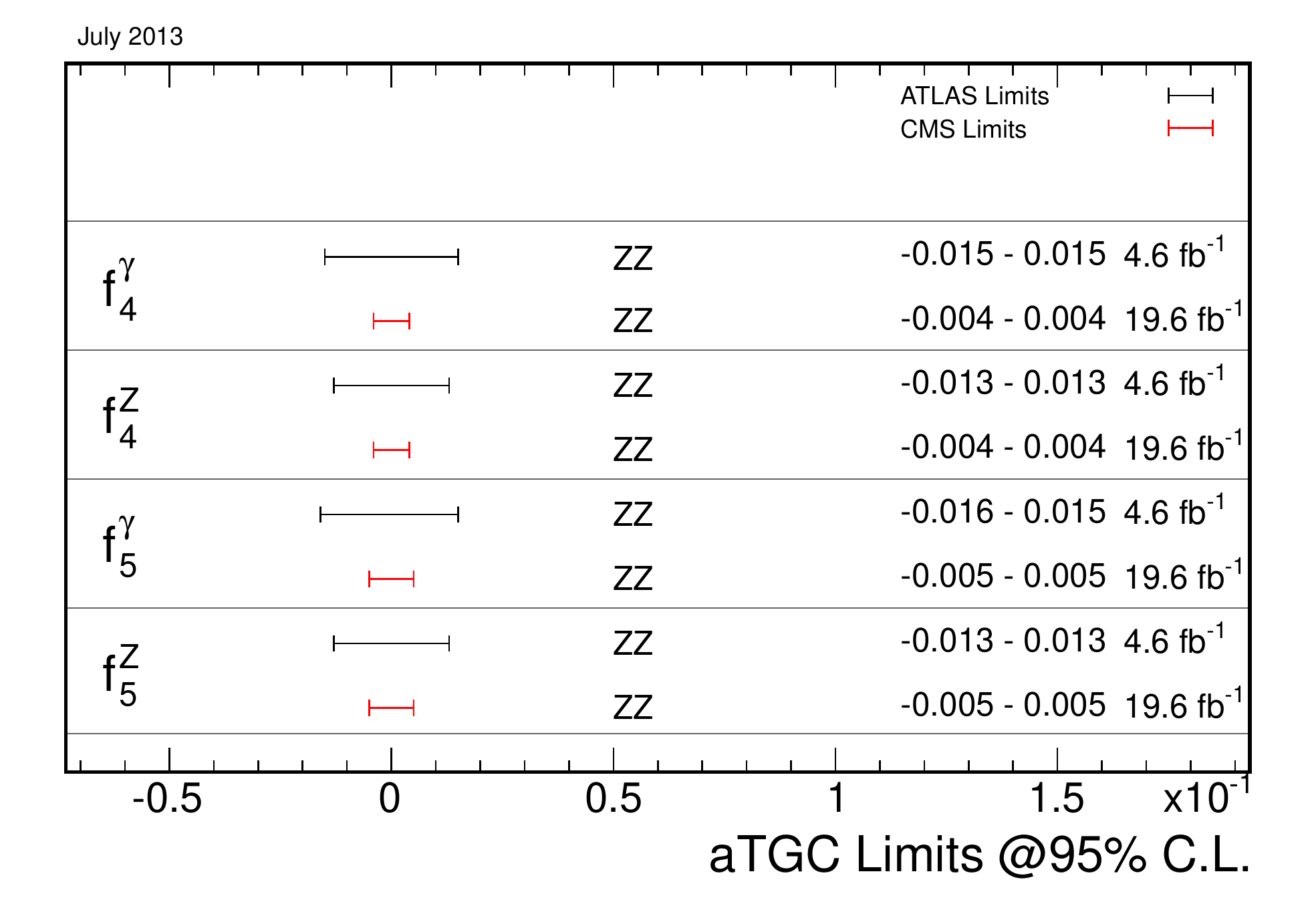}
   \includegraphics[width=0.65\textwidth]{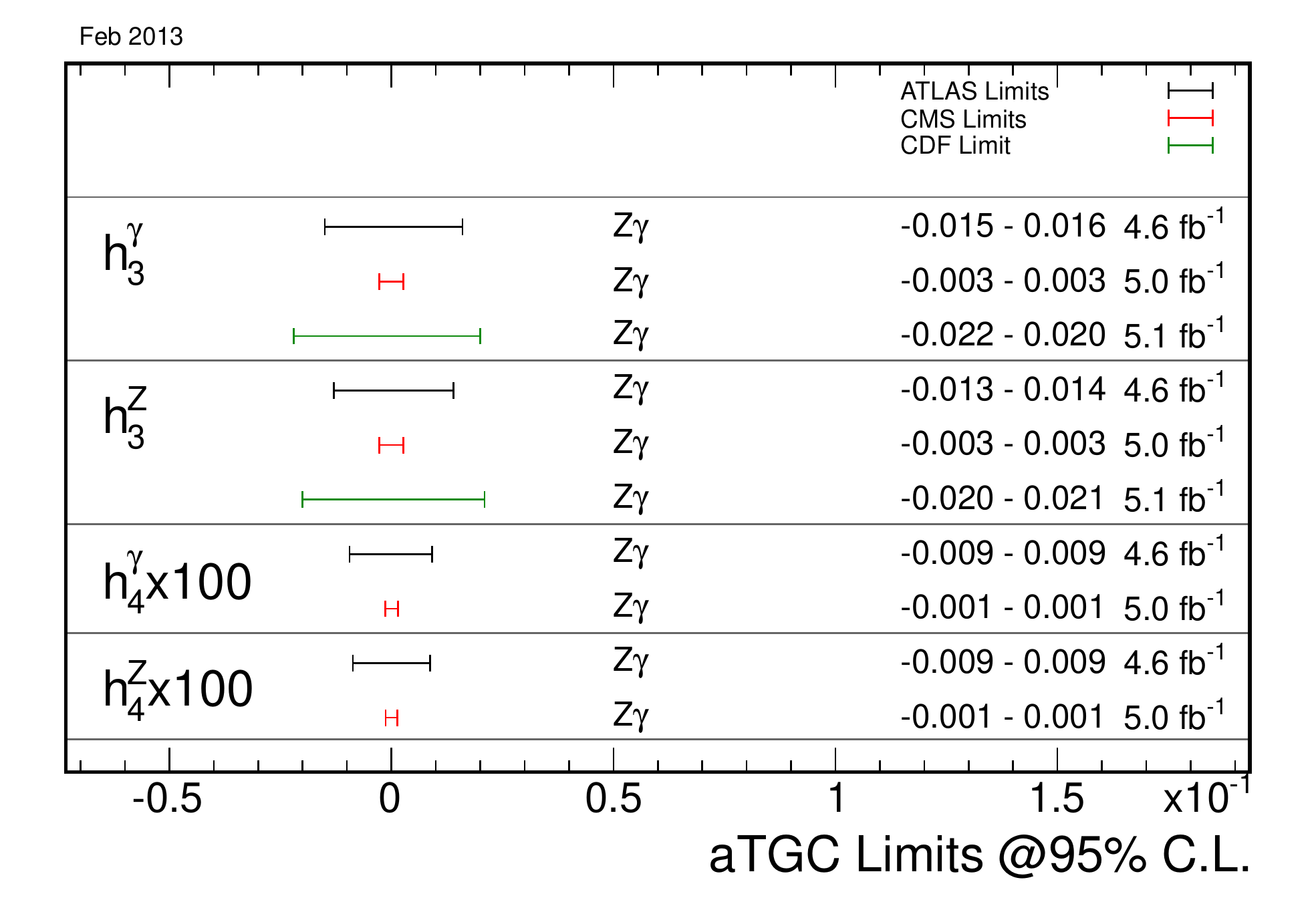}
    \caption{Limits at 95\% C.L. on Z$\gamma$Z, ZZZ (top) and ZZ$\gamma$, ZZ$\gamma$ (bottom) aTGC.}
   \label{fig4}
 \end{center}
\end{figure}

The presence of aTGC would increase diboson production cross sections, in particular at high mass and high $p_{T}$ regions. Experimental measurements made by the ATLAS, CMS, CDF and D\O\ collaborations have found no excess over the SM predictions, leading to limits on charged and neutral aTGC, as shown in Figure~\ref{fig3} and Figure~\ref{fig4} respectively, together with previous LEP results. For charged aTGC, LEP results remain competitive while the sensitivity from LHC is approaching. For neutral aTGC, LHC results dominate.

\section{First Studies On Quartic Gauge Couplings}

CMS has studied the exclusive two-photon production of WW using 5.05~$fb^{-1}$of data at $\sqrt{s}$ = 7~TeV.\cite{Chatrchyan:2013foa} Events are selected by requiring a $\mu^{\pm}e^{\mp}$ vertex with no associated charged tracks, and $p_{T}(\mu^{\pm}e^{\mp}) >$ 30 GeV. Two events are observed in the data, compared to a SM expectation of $2.2\pm0.5$ signal events with $0.84\pm0.13$ background. The significance of the signal is $1.1\sigma$, with a 95\% C.L. upper limit on the SM cross section of 8.4~fb. D\O\ has studied the exclusive two-photon production of WW in events with an electron, a positron and $E^{miss}_{T}$.\cite{Abazov:2013opa} No excess above the background expectation has been found.

A study of the WV$\gamma$, three vector boson production, has also been performed by CMS, using 19.3~$fb^{-1}$ data from $pp$ collision at $\sqrt{s}$ = 8 TeV.\cite{CMS:2013kea} The analysis selects events containing a W boson decaying to electron or muon, a second V (W or Z) boson decaying to two jets, and a photon.
The number of observed events in data is 322, while the estimated background yield is 341.5. This is consistent with the SM predictions, and corresponds to an upper limit of 241~fb at 95\% C.L. for WV$\gamma$ production with photon $p_{T} >$ 10~GeV.

\begin{figure}[hbtp]
 \begin{center}
   \includegraphics[width=0.75\textwidth]{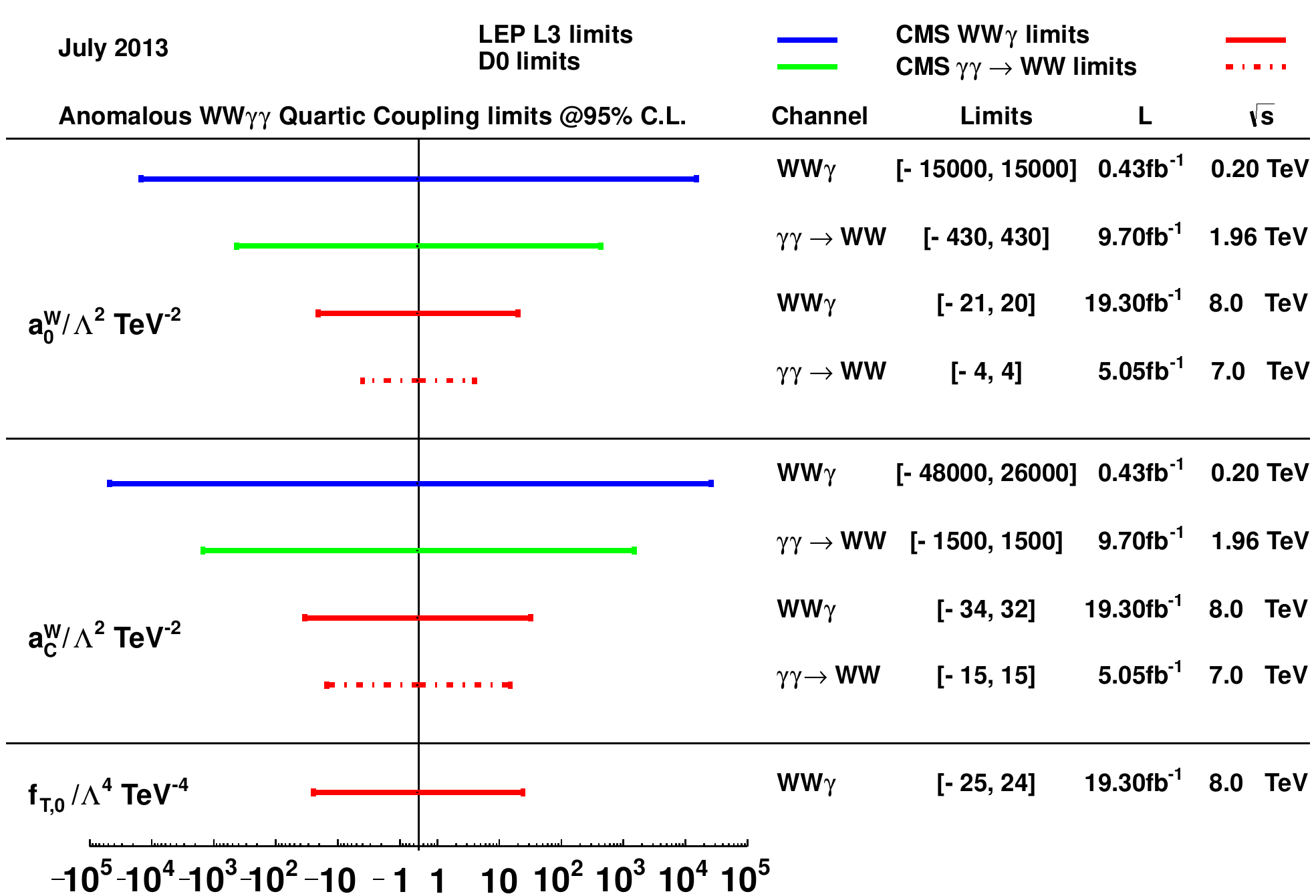}
    \caption{Limits at 95\% C.L. on WW$\gamma\gamma$ aQGC.}
   \label{fig6}
 \end{center}
\end{figure}

The results of the above analyses are studied for deviations from the SM, and used to constrain anomalous quartic gauge couplings (aQGC). The limits on WW$\gamma\gamma$ aQGC set by these measurements, together with previous results from LEP, are shown in Figure~\ref{fig6}. CMS results have significantly surpassed limits from LEP and D\O\ .

\section{Summary}

Latest results of diboson production measurements by ATLAS, CMS, CDF and D\O\ experiments are reviewed.
The measured cross sections are typically consistent with the SM predictions.
The results are used to constrain new physics, by setting limits on anomalous triple gauge couplings. First studies on quartic gauge couplings have started as well.


\begin{thebibliography}{0}    

\bibitem{Aad:2013izg} 
  G.~Aad {\it et al.}  [ATLAS Collaboration],
  Phys.\ Rev.\ D {\bf 87}, 112003 (2013)
  [arXiv:1302.1283 [hep-ex]].

\bibitem{Chatrchyan:2013fya} 
  S.~Chatrchyan {\it et al.}  [CMS Collaboration],
  arXiv:1308.6832 [hep-ex].

\bibitem{Campbell:2010ff} 
  J.~M.~Campbell and R.~K.~Ellis,
  Nucl.\ Phys.\ Proc.\ Suppl.\  {\bf 205-206}, 10 (2010)
  [arXiv:1007.3492 [hep-ph]].

\bibitem{Campbell:2011bn} 
  J.~M.~Campbell, R.~K.~Ellis and C.~Williams,
  JHEP {\bf 1107}, 018 (2011)
  [arXiv:1105.0020 [hep-ph]].

\bibitem{Mangano:2002ea} 
  M.~L.~Mangano, M.~Moretti, F.~Piccinini, R.~Pittau and A.~D.~Polosa,
  JHEP {\bf 0307}, 001 (2003)
  [hep-ph/0206293].

\bibitem{Gleisberg:2008ta} 
  T.~Gleisberg, S.~.Hoeche, F.~Krauss, M.~Schonherr, S.~Schumann, F.~Siegert and J.~Winter,
  JHEP {\bf 0902}, 007 (2009)
  [arXiv:0811.4622 [hep-ph]].

\bibitem{Chatrchyan:2013nda} 
  S.~Chatrchyan {\it et al.}  [CMS Collaboration],
  JHEP {\bf 1310}, 164 (2013)
  [arXiv:1309.1117 [hep-ex]].
    
\bibitem{ATLAS:2012mec} 
  G.~Aad {\it et al.}  [ATLAS Collaboration],
  Phys.\ Rev.\ D {\bf 87}, 112001 (2013)
  [arXiv:1210.2979 [hep-ex]].

\bibitem{Chatrchyan:2013yaa} 
  S.~Chatrchyan {\it et al.}  [CMS Collaboration],
  Eur.\ Phys.\ J.\ C {\bf 73}, 2610 (2013)
  [arXiv:1306.1126 [hep-ex]].
  
\bibitem{Chatrchyan:2013oev} 
  S.~Chatrchyan {\it et al.}  [CMS Collaboration],
  Phys.\ Lett.\ B {\bf 721}, 190 (2013)
  [arXiv:1301.4698 [hep-ex]].

\bibitem{Aad:2012twa} 
  G.~Aad {\it et al.}  [ATLAS Collaboration],
  Eur.\ Phys.\ J.\ C {\bf 72}, 2173 (2012)
  [arXiv:1208.1390 [hep-ex]].

\bibitem{ATLAS:2013fma} 
  [ATLAS Collaboration],
  ATLAS-CONF-2013-021.

\bibitem{CMS:2013qea} 
  CMS Collaboration [CMS Collaboration],
  CMS-PAS-SMP-12-006.

\bibitem{ATLAS:2012dtt} 
  [ATLAS Collaboration],
  ATLAS-CONF-2012-157.

\bibitem{Chatrchyan:2012bd} 
  S.~Chatrchyan {\it et al.}  [CMS Collaboration],
  Eur.\ Phys.\ J.\ C {\bf 73}, 2283 (2013)
  [arXiv:1210.7544 [hep-ex]].

\bibitem{Aad:2012awa} 
  G.~Aad {\it et al.}  [ATLAS Collaboration],
  JHEP {\bf 1303}, 128 (2013)
  [arXiv:1211.6096 [hep-ex]].

\bibitem{ATLAS:2013gma} 
  [ATLAS Collaboration],
  ATLAS-CONF-2013-020.

\bibitem{Chatrchyan:2012sga} 
  S.~Chatrchyan {\it et al.}  [CMS Collaboration],
  JHEP {\bf 1301}, 063 (2013)
  [arXiv:1211.4890 [hep-ex]].

\bibitem{CMS:2013hea} 
  CMS Collaboration [CMS Collaboration],
  CMS-PAS-SMP-13-005.
 
\bibitem{D0:2013rca} 
  V.~M.~Abazov {\it et al.}  [D0 Collaboration],
  Phys.\ Rev.\ D {\bf 88}, 032008 (2013)
  [arXiv:1304.5422 [hep-ex]].
  
 \bibitem{Chatrchyan:2013foa} 
  S.~Chatrchyan {\it et al.}  [CMS Collaboration],
  JHEP {\bf 1307}, 116 (2013)
  [arXiv:1305.5596 [hep-ex]].

\bibitem{Abazov:2013opa} 
  V.~M.~Abazov {\it et al.}  [D0 Collaboration],
  Phys.\ Rev.\ D {\bf 88}, 012005 (2013)
  [arXiv:1305.1258 [hep-ex]].

\bibitem{CMS:2013kea} 
  CMS Collaboration [CMS Collaboration],
  CMS-PAS-SMP-13-009.


\end{thebibliography}

\end{document}